\documentclass[12pt]{article}
 
\usepackage[utf8]{inputenc}
\usepackage[english]{babel}
\usepackage{layout}
\usepackage{verbatim}
\usepackage[top=2.2cm, bottom=2.5cm, left=3cm, right=3cm]{geometry}
\usepackage{graphicx}
\usepackage{listings}
\usepackage{array,multirow,makecell}
\usepackage{mdframed}
\usepackage{hyperref}
\usepackage{parskip}
\usepackage{natbib}
\usepackage{amsmath,amsfonts}
\usepackage{lineno}
\usepackage{setspace}
\usepackage{fancyhdr,color}
\usepackage{etoolbox}
\usepackage[dvipsnames]{xcolor}

\usepackage[toc,page]{appendix}

\usepackage{algorithm}
\usepackage[noend]{algpseudocode}

\usepackage{booktabs}
\usepackage{siunitx}

\title{\Large\textbf{High return level estimates of daily ERA-5 precipitation in Europe estimated using regionalized extreme value distributions}}
\author{Pauline Rivoire$^{1,*}$, Philom\`ene Le Gall$^{2,*}$,\\
Anne-Catherine Favre$^2$, Philippe Naveau$^3$ and Olivia Martius$^{1,4}$}

\definecolor{mygray}{gray}{0.6}

\fancypagestyle{firstpage}
{
    \fancyhf{}
    \chead{\fontsize{10}{12}\leavevmode\selectfont\textcolor{mygray}{Confidential manuscript submitted to \textit{Weather and Climate Extremes}}}
    
}

\begin{document}


\maketitle
\thispagestyle{firstpage}
\vspace*{\fill}
\noindent$^1$ Institute of Geography and Oeschger Centre for Climate Change Research, University of Bern, Switzerland\\
$^2$ Univ. Grenoble Alpes, CNRS, IRD, Grenoble INP, IGE, F-38000 Grenoble, France\\
$^3$ Laboratoire des Sciences du Climat et de l'Environnement, ESTIMR, CNRS-CEA-UVSQ, Gif-sur-Yvette, France\\
$^4$ Mobiliar Lab for Natural Risks, University of Bern, Switzerland\\
\footnotesize  * These authors contributed equally to this work

Correspondence to: Pauline Rivoire (\href{mailto:pauline.rivoire@giub.unibe.ch}{pauline.rivoire@giub.unibe.ch}) and Philom\`ene Le Gall (philomene.le-gall@univ-grenoble-alpes.fr)\\
Orcid IDs: P. Rivoire (\href{ https://orcid.org/
0000-0002-1008-0986}{0000-0002-7231-6210}), P. Le Gall (\href{ https://orcid.org/0000-0002-3758-1764}{0000-0002-3758-1764}), A.-C. Favre (\href{ https://orcid.org/
0000-0002-3314-863X}{0000-0002-3314-863X}), P. Naveau (\href{ https://orcid.org/
0000-0002-7231-6210}{0000-0002-7231-6210}) and O. Martius (\href{https://orcid.org/0000-0002-8645-4702}{0000-0002-8645-4702})

\newpage
\setcounter{page}{1}

\begin{abstract}
Accurate estimation of daily rainfall return levels associated with large return periods is needed for a number of hydrological planning purposes, including protective infrastructure, dams, and retention basins.
This is especially relevant at small spatial scales. The ERA-5 reanalysis product provides seasonal daily precipitation over Europe on a 0.25$^\circ \times$0.25$^\circ$ grid (about 27 $\times$ 27 [km]). 
This translates more than 20,000 land grid points and leads to models with a large number of parameters when estimating return levels. 
To bypass this abundance of parameters, we build on the regional frequency analysis (RFA), a well-known strategy in statistical hydrology. 
This approach consists in identifying \emph{homogeneous} regions, by gathering locations with similar distributions of extremes up to a normalizing factor and developing sparse regional models. 
In particular, we propose a step-by-step blueprint that leverages a recently developed and fast clustering algorithm to infer return level estimates over large spatial domains.
This enables us to produce maps of return level estimates of ERA-5 reanalysis daily precipitation over continental Europe for various return periods and seasons. We discuss limitations and practical challenges and also provide a git hub repository.
We show that a relatively parsimonious model with only a spatially varying scale parameter can compete well against statistical models of higher complexity.
\end{abstract}


\newpage
 
\section{Introduction}\label{sec: intro}
Heavy rainfall can cause natural hazards such as landslides, avalanches and floods  \citep[e.g., see][]{ipcc2013, EEA2018}. Such hazards can cause casualties and damages, with direct and indirect economic impacts \citep{RE2018, Prahl2018}.
To design protective infrastructure, for instance, a dam, one needs to know the frequency of a given intensity of precipitation \citep{Madsen2014}. The return-period of an event is the duration during which the event occurs once, \emph{on average} \citep[see e.g.][]{cooley2013}. Symmetrically, for a given duration, say 100 years, the 100-year return level is defined as the level that is exceeded once every 100 years, \emph{on average}. Given a dataset (e.g. observation or reanalysis), time series are finite and observing an event exactly once in 100 years does not make an event the 100-year return level. One therefore needs a statistical model to predict the intensity of such events, and even unobserved events.

The aim of this paper is to provide return levels for large return periods over Europe. The station coverage being quite heterogeneous over Europe \citep{Cornes2018} therefore, the use of gridded datasets is appropriate. Various types of gridded precipitation data sets are available \citep[e.g., see][for an overview]{Sun2018}. Precipitation gridded data can be derived from ground observations, satellite observations, combinations of ground observations and satellite observations and short-term numerical weather forecasts in reanalysis datasets. In reanalyses, past observations are assimilated in numerical weather forecast models to reconstruct past weather. The main advantage of this type of data set is its regular spatial and temporal coverage. Reanalyses also ensure consistency of the precipitation data with the atmospheric conditions, which is a valuable characteristic for weather and climate process studies. Precipitation in this study is extracted from the ERA-5 reanalysis data set \citep{CopernicusERA5, Hersbach2020}. We study daily precipitation over continental Europe. The region of interest covers more than 20,000 grid points over European land.

Extreme value theory (EVT) provides an asymptotic framework to model the distribution of extremes such as heavy precipitation. Two classical approaches for extreme modelling are the generalized extreme value (GEV) and the generalized Pareto distribution (GPD).
The GEV \citep{jenkinson1955} aims at modelling maxima over large blocks \citep[for instance, a year in][]{poschlod2021}.  The  GPD \citep[see e.g. ][ and Section \ref{sec: EGPD model}]{pickands1975} enables the modeling of exceedances over a given threshold \citep[for instance, the $98$-th quantile in][]{car17}. However, these two approaches only model extremes and our goal is to provide return levels in the full rainfall intensity range. We therefore need a class of distribution that can model the whole spectrum of precipitation intensities. \cite{carreau09}, \cite{papastathopoulos2013}, \cite{nav16} and \cite{Stein20} introduced distributions that model the whole spectrum of rainfall intensities. The methods model the upper tail with a Pareto distribution. Various types of transfer functions then fit the bulk and lower tail distribution. \cite{tencaliec20} defined a flexible version of the extended generalized Pareto distribution (EGPD) and \cite{Rivoire2021} used it to fit the positive daily precipitation of ERA-5. The transfer function is estimated using Bernstein polynomials which bring flexibility to the transfer function estimation but require a large number of parameters \citep[for example 30 for each grid point in][]{Rivoire2021}. In this paper, we simply use a monomial transfer function with a single flexibility parameter, see Section \ref{sec: EGPD model} for more details.

 \cite{pos2021} fitted  GEV distributions at each grid point of ERA-5 in Bavaria (Germany) to estimate  10-year precipitation return levels. However, extending this pointwise analysis of precipitation across Europe is quite onerous. Fitting a GEV and computing return levels for each grid point requires the estimation of more than $3\times 20,000$ parameters (location, scale and shape parameters). In addition, estimates of the shape parameter at a specific location are quite sensitive to the length of the time series \citep[e.g., see][]{zhang12,malek14,jalbert17}. Therefore reducing the dimensionality of the fitted parameters is of great practical importance. 
In contrast to this local approach, \cite{sang2009} and \cite{naveau2014} assumed the shape parameter to be constant over the area of interest (Cape Floristic Region in South Africa and Switzerland, respectively). However, Europe is much larger than these areas, and the diverse climate and complex orography \citep{beck2018, ESOTC2020, ECMWFclimato} strongly influence the spatial distribution of precipitation \citep[e.g., see][]{evin16, marra2021}. The method used for dimensionality reduction should preserve the diverse spatial patterns of precipitation over Europe. In this paper, we therefore consider an intermediate approach in which the shape parameter is common between grid points within homogeneous regions.

  The regional frequency analysis (RFA), a concept from hydrology, attempts to build these homogeneous regions which consist of grid points with similar precipitation distributions \citep{dal60,hos05}. In a homogeneous region, distributions are all equal to a common regional distribution up to a normalizing factor. In particular, their extreme behaviour should be analogous. Clustering grid points in homogeneous regions reduces the dimensionality of large precipitation data sets while preserving the spatial patterns. We use the definition of homogeneous distributions proposed by \cite{st03} and \cite{hos05}: 
given a region of interest, say $\mathcal{R}$ (here Europe), a homogeneous cluster ($\mathcal{C}$)  is defined as a sub-region where all spatial points $s$, have the same  marginal distribution up to normalization:
\begin{equation}\label{eq: defhomo} 
\mathcal{C} = \left\{ s \in \mathcal{R} : \;\;  Q_s=\lambda(s) \times  q \; \right\}, 
\end{equation}
where $Q_s$ is the quantile function at site $s$, the positive scalar $\lambda(s)$  varies in space, and   $q$ represents  a positively-valued  and dimensionless quantile function (common to every site in the cluster). As a consequence  rescaled quantiles within a homogeneous cluster do not depend on localization $s$.

Several methods allow regions to be delineated as in Eq. \eqref{eq: defhomo}.
They often require climate and/or geographical covariates \citep[see e.g.][for recent work]{fawad18, for18} and work in three steps: i) selecting explanatory covariates, ii) grouping sites with similar covariates, and iii) testing the homogeneity of the groups obtained.
Covariates are selected for their ability to explain the precipitation distribution \citep{evin16, oua08}.  For instance, \cite{dar21} selected explanatory covariates by applying a principal component analysis to available geographical and climate data. They found that longitude, latitude, elevation and seasonality of events explained most hourly precipitation in the UK. With these methods, choosing covariates is an essential step that requires expert knowledge. Moreover, covariate data must be available and may be complicated to transfer across regions with different climate characteristics. For example, the covariates that best describe precipitation may be different between the UK and Italy. 
To check the homogeneity of covariate-based groups, \cite{hos05} proposed tests to examine the  validity of the model corresponding to Eq. \eqref{eq: defhomo}. The tests rely on two components:  the moments that characterize the precipitation distribution, and the distributional assumption \citep[Kappa-distributed, see e.g.][]{hosking1994}. The tests consist of measuring the dispersion of some estimated L-moments (for all sites in the region) around a theoretical regional value of L-moments. To compute the theoretical value, \cite{hos05} assume that the precipitation follows a Kappa distribution. This distributional assumption is not necessarily satisfied in practice. 
To bypass the selection of covariates, \cite{saf2009} and \cite{legall21} proposed methodologies using precipitation data only. They started from the hypothesis that the distributions are partially characterized by their probability weighted moments \citep[PWM,][]{green79}. 

\cite{legall21} recently proposed a PWM-based algorithm to identify homogeneous spatial clusters of extreme precipitation and applied the algorithm to Swiss daily precipitation observations. The algorithm provided spatially coherent regions without using any geographical covariate. In this paper, we apply the clustering algorithm from \cite{legall21} to ERA-5 daily precipitation from all European land areas to group grid points with similar upper tails.

When clusters are delineated, information from all homogeneous grid points can be pooled to accurately estimate the EVT distribution parameters. For the regional distribution, we use an EGPD with three parameters see Section \ref{sec: EGPD model}. Only the scale parameter can vary within a homogeneous cluster. The flexibility and shape parameters are constant over the cluster. 
In a nutshell, the regional approach allows us to go from a model with $3\times20,000$ parameters to a model with $2 \times n_{\textrm{clusters}} + 20,000$ parameters, $n_{\textrm{clusters}}$ being the number of clusters. We also compare the performance of this regional approach to the performance of a more flexible distribution where the flexibility parameter can vary between sites of the same homogeneous cluster.

This study is the first to provide ERA-5 return levels, which, to our knowledge, have never been provided for the whole of Europe. Second,  RFA is traditionally applied to smaller areas such as countries \citep[see e.g.][ for RFA on the UK and Switzerland]{fowler2003,  evin16}.

Section \ref{sec: data} introduces the precipitation data set and Section \ref{sec: method} describes the methods for the non-parametric clustering algorithm, the regional fitting and its assessment.
The homogeneous regions, the assessment of the regional fitting and the corresponding 10, 50 and 100-year return levels are presented in the results section, Section \ref{sec: results}. We discuss our results and compare our clusters to the regions obtained by national-scale studies in Section \ref{sec: discussion}. We draw conclusions in Section \ref{sec: conclusion}.

\section{Data}\label{sec: data}

We use ERA-5 daily precipitation with 0.25$^{\circ}$ spatial resolution. ERA-5 is the latest global reanalysis data set provided by the European Center for Medium-Range Weather Forecasts \citep{CopernicusERA5, Hersbach2020}. Precipitation is provided with hourly resolution forecasts that we aggregate to daily precipitation. We study ERA-5 precipitation for the period 1979--2018 in Europe over land, which is a region in which the data set performs well \citep{Rivoire2021}. Because practical applications are mainly restricted to the continent, we do not include precipitation data over the oceans. We conduct a seasonal analysis to ensure the stationarity of the time series. We consider the daily positive precipitation for each season. Days are considered as wet when precipitation exceeds $1$ mm \citep{Maraun2013}.

\section{Methods}\label{sec: method}

Here, we introduce the two stages of RFA: i) identify homogeneous regions (Sections \ref{sec: omega} and \ref{subsec: clustalgo}) and ii)  use data from all grid points in the same region to model rainfall intensities (Section \ref{sec: EGPD model}). We also introduce the evaluation tools we used to assess the fitted distributions (Section \ref{sec: metho_assess}).

 \subsection{A scale-invariant ratio of PWM}\label{sec: omega}
 Following the notations of \cite{legall21}, we denote $\alpha_i (Z)$ the $i$-th PWM of the positive $F$-distributed random variable $Z$
 \begin{equation*}\label{eq: PWM}
     \alpha_i (Z) = \mathbf{E}\left[Z F(Z)^i\right].
 \end{equation*}
 When self-evident, it is denoted simply as $\alpha_i$. The first three moments are used to compute the scale-invariant ratio 
 \begin{equation}\label{eq: def omega}
    \omega = \dfrac{3 \alpha_2 - 2 \alpha_1}{2\alpha_1 - \alpha_0}.
\end{equation}

\cite{legall21} showed that $\omega$ can be seen as a ratio of two distances derived from norms. 
 
Let $i$ and $j$ be two grid point locations, and $Y_j$ and $Y_j$ their two associated time series of seasonal positive precipitation. To spatially cluster daily rainfall, we need to compute a dissimilarity measure between two positive time series. Here, we use the $\omega$-based distance defined by 
 \begin{equation}\label{eq: omega_dist}
     \hat{d}_{ij} = \left|\widehat{\omega \left(Y_i\right)} - \widehat{\omega\left(Y_j\right)} \right|,
 \end{equation}
where $\widehat{\omega \left(Y_i\right)}$ is the estimate of $\omega(Y_i)$.
We use this distance for two reasons. First, because the distance is based on PWM, it enables comparison of empirical distribution shapes, including heavy-tailed ones, without fitting a parametric distribution. Second, the key property of $\omega$ is its scale-invariance. For any precipitation variables $Y_1, Y_2$ in a homogeneous region, see Eq. \eqref{eq: defhomo}, 
\begin{equation*}
    \omega(Y_1) = \omega(Y_2).
\end{equation*}
The ratio $\omega$ can be interpreted as the heaviness of the tail within the mathematical framework of EVT.  In the block maxima or peak-over-threshold approaches, $\omega$ only depends on the shape parameter. In the  EGPD approach, $\omega$ depends on the shape and the flexibility parameter, see Section \ref{sec: EGPD model}.
The distance between two grid points with homogeneous distributions should be close to zero. 
 The clustering algorithm gather sites with similar $\omega$ estimates.

 \subsection{Clustering algorithm: partitioning around medoids (PAM)}\label{subsec: clustalgo}
Grouping close $\omega$ estimates is an unsupervised learning problem: we gather unclassified points that have common characteristics (here, their $\omega$ value). The grouping of estimates into clusters is based on geometric considerations: estimates are grouped if they are close to each other in the space of variables (here the axis of reals).

 Several clustering methods are available \citep{kauf90, murty1999, schubert2021}, most classic ones fall in two categories: partitioning or hierarchical methods. 
The $k$-medoids, or partitioning around medoids (PAM), and $k$-means are iterative algorithms that belong to the first group. They both require the final number of clusters $k$ as input. The PAM algorithm is preferred to the $k$-means because of its ease of interpretation. Indeed, centres of the $k$-means clusters are barycentres and therefore virtual points whereas the centres of the PAM clusters are actual points of the dataset \citep[see e.g.][]{jain1999, ber13}. 
 For each of these methods, the choice of the dissimilarity measure is paramount. We work with the absolute difference as a distance, also called the Manhattan distance, as recommended in \cite{ber13, bad15}, see Eq. \eqref{eq: omega_dist}.

The centre of each cluster is the grid point with the smallest dissimilarity to all other grid points in the cluster and is called the medoid.
Each non-medoid point of the data-set is associated with its closest medoid.
Generally speaking, PAM converges to an ensemble of medoids and clusters that is a local minimum of the total cost, see Eq. (9) in \cite{legall21}. 

To solve this optimization problem, PAM starts by selecting $k$ initial medoids, here in a deterministic way.
The first medoid is the medoid of the partition for one cluster: the most centrally located point.
The set of $k$ medoids is then completed by adding the medoids of partitions with an increasing number of clusters one by one until $k$ is reached.

The second step consists of testing every swap possible between a medoid and any point non-medoid in the whole data set. If the total cost function \citep[see e.g. ][]{legall21} decreases, then the point is kept as medoid. Clusters are then updated with respect to their new medoids. When no swap decreases the total cost, then the algorithm stops. \\
The computational cost of these two steps increases with the size of the data set and the number of clusters. Because ERA-5 provides data for about 20,000 grid points in European lands, we use a faster version \citep{reynolds2006, schubert2021} of the original algorithm. This variation removes some redundant computations in the swap step. 

To measure the strength of the link between a point and its cluster, \cite{rou87} introduced the silhouette score.
 The silhouette score for grid point $i$ that belongs to the cluster $k$ is defined as 
\begin{equation}\label{eq: sil}
    1-\left(\dfrac{d_{ik}}{\delta_{i,-k}}\right)
\end{equation}
 where $d_{ik}$ is the average intra-cluster dissimilarity between grid point $i$ and all other grid points in cluster $k$, and
 $\delta_{i,-k}$ the smallest of the $k-1$ average distance between site $i$ and all other sites associated with a cluster different from $k$. 
 When a grid point $i$ is well classified, the intra-cluster average distance is significantly smaller than the distance between clusters. Its silhouette score is then close to 1. By contrast, a silhouette score close to -1 indicates a poorly classified grid point that should be in another cluster. Eventually, a grid point that is not significantly closer to points in the cluster than to other points has a silhouette score close to 0. In other words, it is not strongly linked to any cluster.
 
 Finding the optimal number of clusters in a data set is a tricky task \citep{ sugar2003, pan13}. Numerous criteria that aim at identifying tight and well-separated clusters exist \citep{halk02, desgraupes2013}. 
 We compute five of them \citep[silhouette, Dunn, Davies Bouldin, Xie Beni, S\_Dbw, see e.g.][]{halk02, desgraupes2013} to determine the optimal number of clusters, between two and ten. These criteria are based on different distances and provide a different optimal number of clusters.
We therefore choose the number of clusters subjectively. We visually compare the maps of the partitions for numbers of clusters. We compromise between a large number of clusters and a partition that is not fragmented.

\subsection{Regional fitting}\label{sec: EGPD model}

To model the entire precipitation distribution, \cite{nav16} and \cite{tencaliec20} proposed a simple scheme to build a flexible distribution by writing
\begin{equation*}
    F(z) = G(H_{\sigma,\xi} (z))
\end{equation*}
where the flexibility function $G$ can be any cumulative distribution function such that there exists $\kappa >0$ such that $\dfrac{G(u)}{u^\kappa}$ and $\dfrac{1-G(1-u)}{u}$ have finite limits when $u$ goes to zero. These constraints ensure that $F$ follows  EVT for very low and high precipitation accumulations.  
 Here, we use $G_{\kappa}(u) = u^\kappa, \quad  \kappa >0$, as flexibility function. Although simple, $G$ is sufficiently flexible to model daily rainfall distributions while maintaining parsimony in the model \citep{evin16}. 
 
 We fit the parameters to different levels of regionalization, from $\sigma$, $\xi$ and $\kappa$ computed individually at every grid point to $\sigma$ computed individually and $\kappa$ and $\xi$ being common between grid points in a homogeneous region (see Table \ref{tab: EGPD_models}).

PWM can  quickly be estimated non-parametrically and used for estimation of EGPD parameters \citep[see Appendix of ][]{nav16}. Estimates of local parameters are taken as initial values for the iterative estimation of regional or semiregional parameters, see Algorithm \eqref{alg: regEGPDfitIter}. 
 
 The quantile with probability $p$ can be computed using the explicit formula
\begin{equation}
   y_p =  F^{-1}(p) = \dfrac{\sigma}{\xi} \left[ \left\{1-G^{-1}(p)\right\} -1 \right], \quad \text{if } \xi >0,
\end{equation}
$0<p<1$. The return level associated with return period of $T$ years is $y_p$ for $p=\frac{1}{T\times n_\textrm{wds}}$, where $n_\textrm{wds}$ is the number of wet days per season. We use the mean of the number of wet days per season during the period under study as an approximation for $n_\textrm{wds}$.

For every grid point, we assume that the random variable modelling daily positive precipitation is independent and identically distributed. However, precipitation events can last for several consecutive days \citep{buritica2021}. To ensure independence in a time series of wet days, we extract one wet day out of three to fit the EGPD models. Despite the climate change, \cite{Donat2014} did not detect any clear trend in the whole precipitation distribution over the period of interest. The absence of a trend and the seasonal analysis are necessary to ensure identical distribution.

\subsection{ Assessment of the fitting} \label{sec: metho_assess}
We evaluate the goodness-of-fit with standard statistical tools focusing on accuracy, flexibility of estimation, and rewarding of the parsimony (smaller number of parameters).

First, quantile-quantile plots (QQ-plots) provide visual information on the proximity between two distributions. For selected grid points, we present QQ-plots, contrasting the empirical quantiles with the quantiles parametrically estimated with the local, semiregional and regional fits, and EGPD with Bernstein flexibility function (see Table \ref{tab: EGPD_models}).

We assess the agreement between the fitting and the empirical distribution with the Anderson-Darling test \citep[see e.g.][]{anderson1952, scholz87}. To ensure independence between the empirical and fitted distribution at a given grid point, we use a third of the positive precipitation time series that was not used in the fitting process as empirical data. Table \ref{tab:AD_results} summarizes the results of the Anderson–Darling test over Europe for the regional, the semiregional, and the local fittings. To ensure spatial independence, we perform the tests for 1/8th of the grid points, randomly chosen. This way we avoid repetition of information between neighbouring grid points. 

To evaluate the goodness-of-fit, we compute the Akaike information criterion (AIC) \citep{akaike1987} for the local, the semiregional, and the regional models. This criterion combines a measure of the goodness of fit (log-likelihood) with the parsimony andsparsity of the model. The AIC has to be minimized. A smaller number of fitted parameters is a bonus for the model because this reduces  the risk of overfitting. For example, the local model requires the estimation of about $3 \times 20,000  $ parameters, whereas the regional model only needs the estimation of about $20,000 + (\text{number of clusters})\times2$ parameters.

\section{Results}\label{sec: results}

\subsection{Partition of ERA-5 over Europe}\label{sec: clust_res}
We apply the clustering algorithm introduced in Section \ref{subsec: clustalgo} to ERA-5 positive daily precipitation for each season independently.

\begin{figure}[h!]
    \centering
    \includegraphics[width=\textwidth]{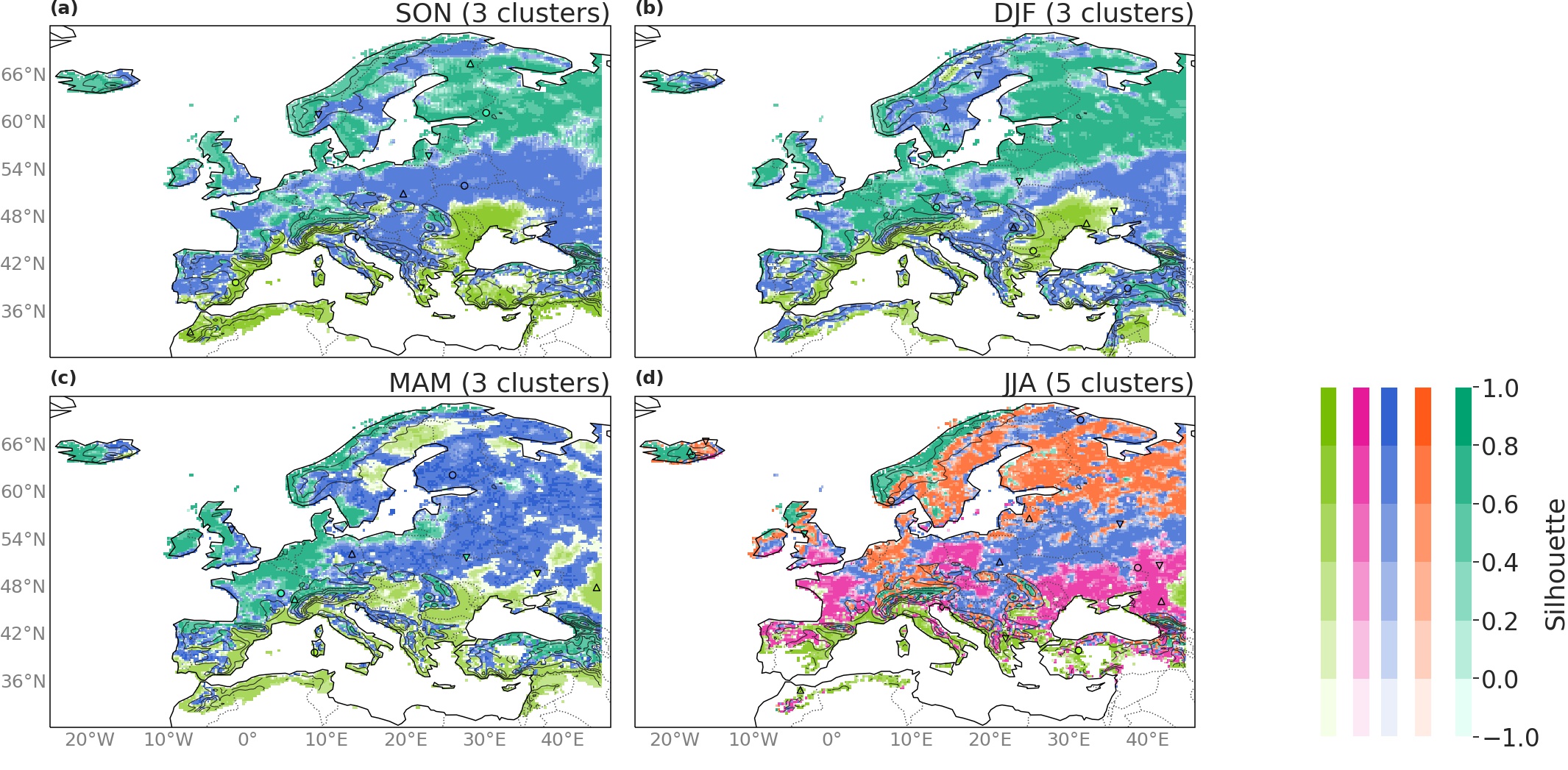}
    \caption{Partition with the PAM algorithm applied on ERA-5 daily positive precipitation over Europe for all seasons. Each cluster is identified by a colour. The shades of colour indicate the silhouette coefficient at every grid point. Intense hues indicate a strong association with the cluster. The black lines are 500 [m] altitude isolines of the surface topography in ERA-5. Within a cluster, the circle indicates the location of the medoid, and the triangle pointing up (resp. down) indicates the grid point with the highest (resp. lowest) silhouette coefficient.}
    \label{fig:partition_PAM}
\end{figure}

The optimal number of clusters is three for September-October-November (SON), December-January-February (DJF), and March-April-May (MAM), and five for June-July-August (JJA, see Section \ref{sec: discussion} for a discussion about this number). Figure \ref{fig:partition_PAM} shows these partitions. The shade of colour indicates the silhouette coefficient of the grid points; light colours indicate low silhouette coefficients and therefore a weak association with the cluster.
There are very few isolated grid points. For all the seasons, the borders between clusters follow the orography, for example in the Alps, the Carpathians, and the UK. This orographic link is present in all seasons. Hence, the ratio $\omega$ captures spatial structures associated with physical features such as orography without requiring additional covariates such as longitude, latitude, or elevation. Silhouette scores are lowest at the borders between clusters, and downward-pointing triangles, which indicate grid points with low and minimum silhouette coefficients, are often located in transition zones between clusters (Fig. \ref{fig:partition_PAM}).

\subsection{Assessment of the fitting}\label{sec: assess_fit_res}

The fitting models are assessed with the Anderson--Darling test, the AIC criterion and QQ-plots.

The Anderson--Darling test indicates similar performance for the fitting of the regional and local EGPD models; see Table \ref{tab: EGPD_models}. The null hypothesis is that the fitted and the empirical distribution are the same. Table \ref{tab:AD_results} displays the  nonrejection rates of the null hypothesis for the Anderson--Darling test for each season and model across the entire domain.
The null hypothesis is not rejected for 87\% of the grid points in JJA and 91\% in SON for the local fit. For the regional fit, it is not rejected for 84\% of grid points in JJA and 88\% in SON, DJF, and MAM. The nonrejection rate for the semiregional fitting is very similar to that the regional fitting. The percentage is lower for the local fit than for the regional fit in all seasons. Nonetheless, the difference between local and regional is smaller than 3\% in all seasons. For all seasons and all fittings, the nonrejection rate indicates good performance of the model, the perfect nonrejection rate being 95\% on a test with a confidence level of 5\%. 

The variability of meteorological processes tends to increase with the altitudinal gradient. Around complex topography, local-scale variations in precipitation may occur. Precipitation distributions might differ substantially between grid points, even within a homogeneous region, and the quality of the regional fit might decrease. We therefore distinguished the rejection rate of Anderson--Darling test between grid points below and above 1000 meters above sea level. We did not find any significant difference in the rejection rate of the Anderson--Darling test between grid points at low and high altitudes (not shown). Moreover, the goodness of the classification in the clustering procedure might impact the accuracy of the fit. At a grid point with a poor connection to its cluster, the regional value of $\xi$ (and $\kappa$) might not be accurate and the distribution fitted regionally might be significantly different from the empirical distribution. We distinguished the Anderson--Darling test between grid points with a silhouette greater or lower than 0.2. Here too, we observe no significant difference between grid points with low and high silhouettes, for either the local or regional fits (not shown). 
Even if the local model is the most adaptable, the regional model seems to be sufficiently flexible to (i) take into account the local-scale variations caused by complex topography and (ii) compensate for the regionalization of two parameters out of three.

The AIC criterion summarizes the information contained in the likelihood and penalizes the number of parameters. It should be as low as possible.
The AIC is much lower for the regional model than for the semiregional and local models independent of the season (see Table \ref{tab:AIC_results}). AIC values across all grid points vary between $-115,106$ in JJA and $-107,250$ in DJF for the regional fitting, between $-79,400$ in JJA and $-67,614$ in DJF for the semiregional fitting and between $-43,704$ in JJA and $-27,984$ in SON for the local fitting. These AIC values highlight the trade-off between parsimony and goodness of fit of the regional fitting. Having only one EGP parameter varying within one cluster in the regional model substantially reduces the AIC.

 Figure \ref{fig:qqplot} displays the QQ plots for cluster medoid, cluster minimum, and cluster maximum silhouette coefficient in each cluster in SON (see partition in Fig. \ref{fig:partition_PAM}(a)). For the most centrally located grid point, the medoid, all the fittings perform similarly well. One exception is the upper tail in the northern and southern clusters, which is slightly overestimated with the regional fitting compared to the local one. For the grid point with a minimum silhouette, the regional and semiregional fit have a similar performance to the local one or even outperform them in the southern cluster. In the intermediate and southern clusters, for the semiregional and regional fittings only the most extreme precipitation is overestimated compared to the local fit. For the grid point with the maximum silhouette, the regional and semiregional fits outperform the local fit for the whole distribution. Extremes are well captured with the regional method, except in the northern cluster, where the highest precipitation is overestimated for all the fittings.
The semiregional and regional fittings seem to significantly improve the quality of estimation for the best-classified points. The semiregional and regional fittings have similar performances. We also compared with the local Bernstein fitting; see Table \ref{tab: EGPD_models}. Its performance is similar to the semiregional and regional fittings except in the southern cluster.

\subsection{Return levels}\label{sec: regfit_RL}
\begin{figure}[h!]
    \centering
    \includegraphics[width=12cm]{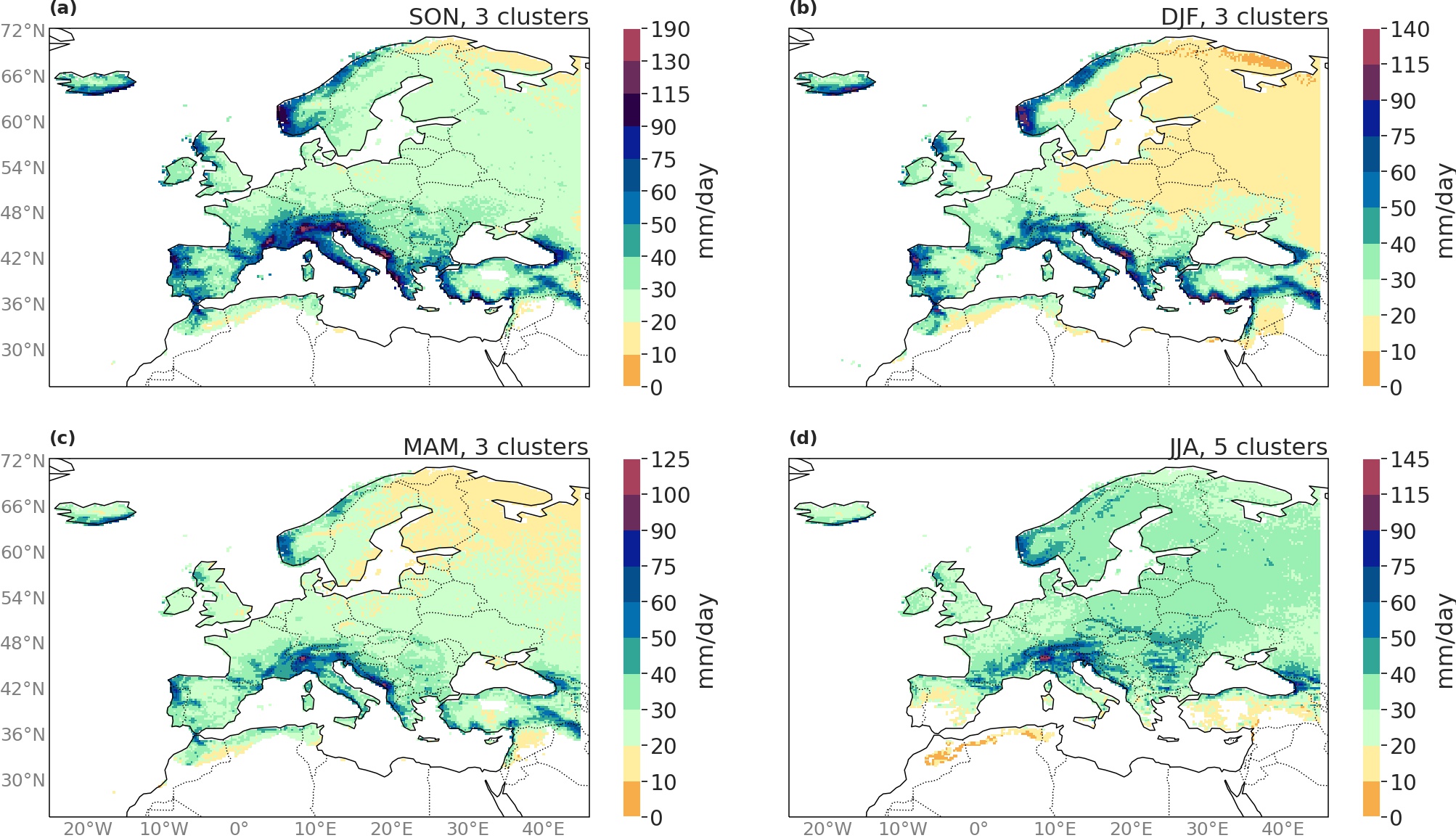}
    \caption{10-year return levels computed with the regional fitting, see Table \ref{tab: EGPD_models}.}
    \label{fig:RL_10}
\end{figure}

\begin{figure}[h!]
    \centering
    \includegraphics[width=12cm]{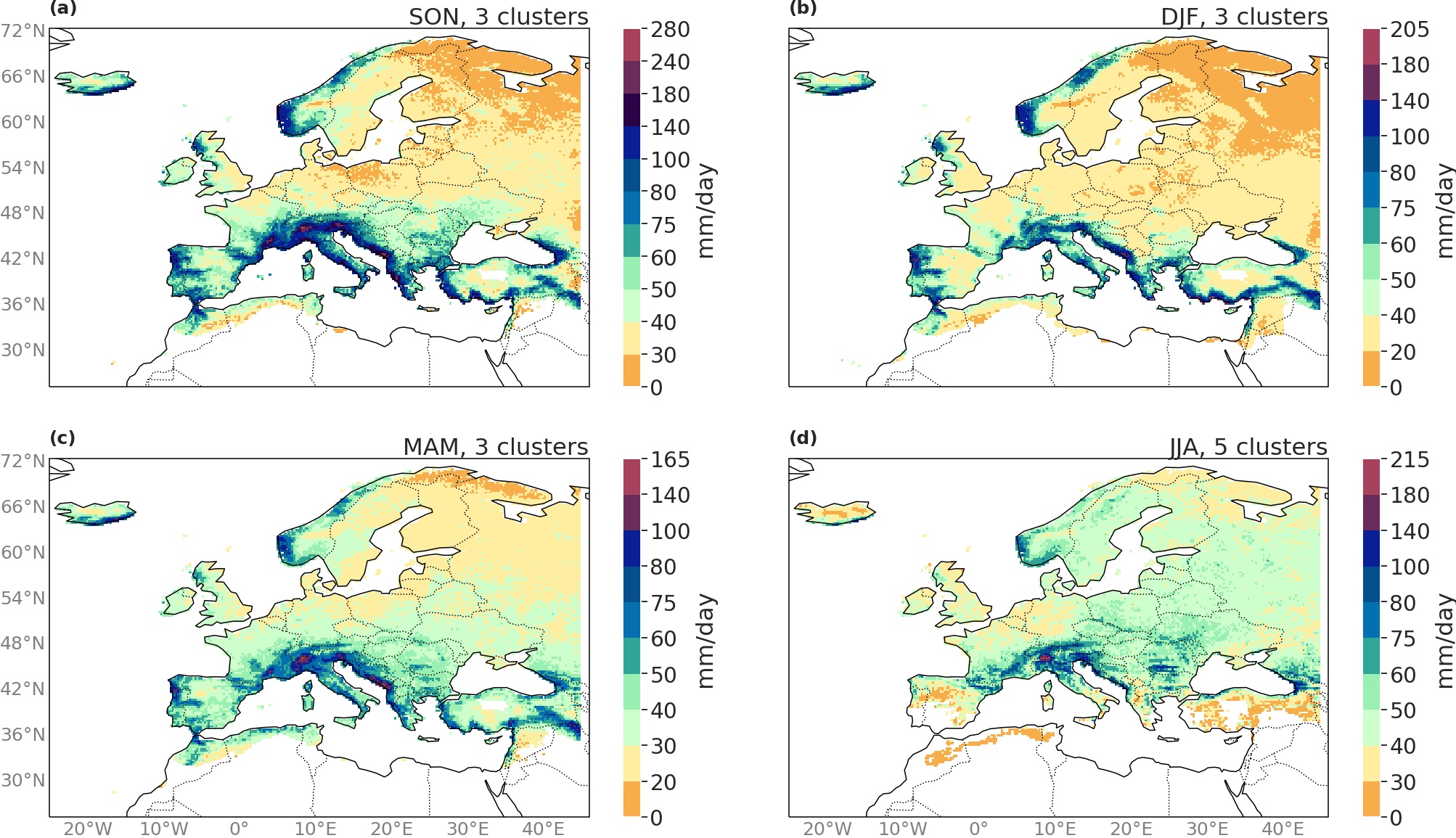}
    \caption{50-year return levels computed with the regional fitting, see Table \ref{tab: EGPD_models}.}
    \label{fig:RL_50}
\end{figure}

\begin{figure}[h!]
    \centering
    \includegraphics[width=12cm]{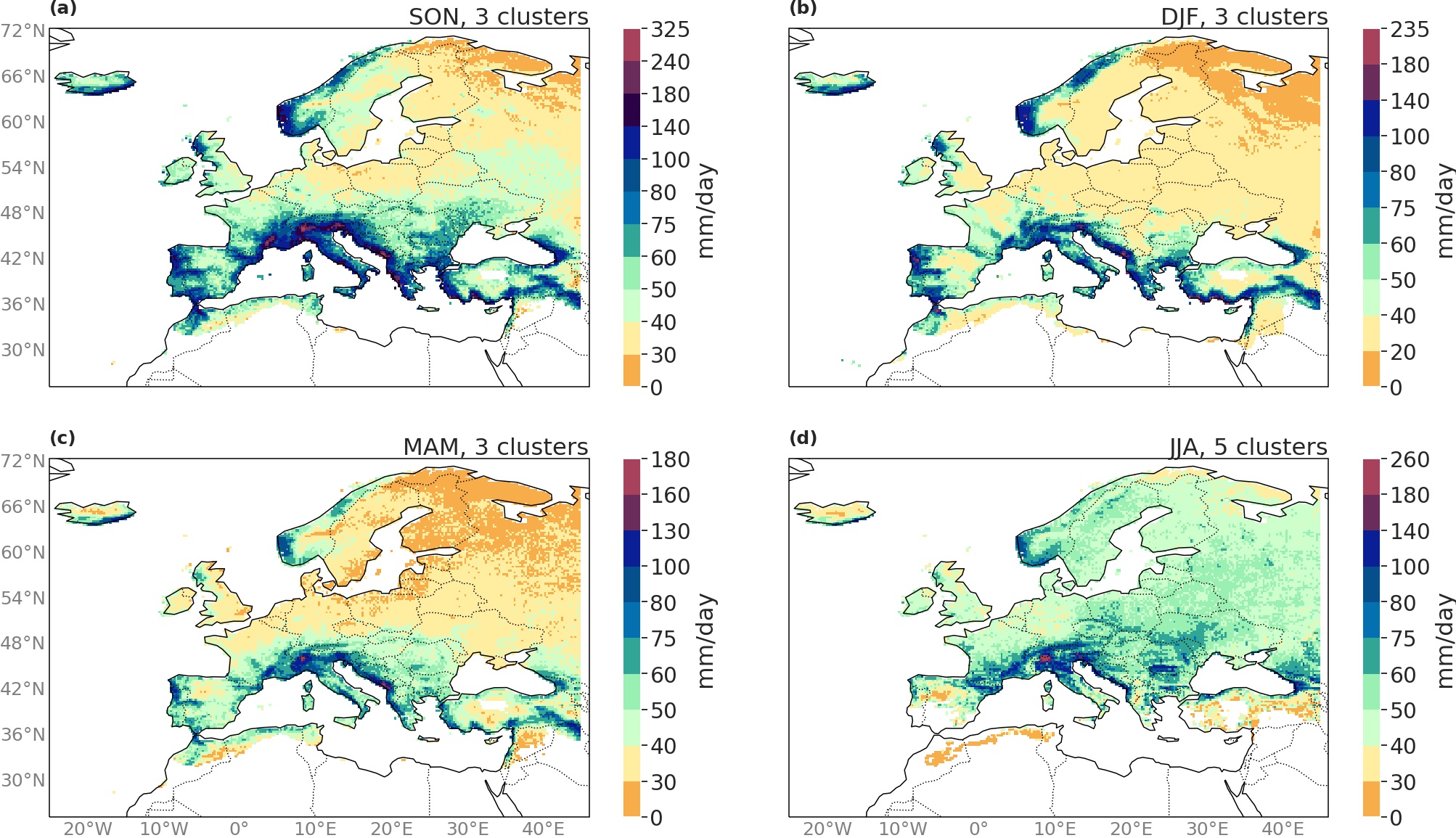}
    \caption{100-year return levels computed with the regional fitting, see Table \ref{tab: EGPD_models}.}
    \label{fig:RL_100}
\end{figure}

The estimate of the return levels is spatially smooth despite the regionalization of two out of three parameters in the EGPD. Figures \ref{fig:RL_10}, \ref{fig:RL_50} and \ref{fig:RL_100} show the 10-, 50-, and 100-year return levels for all seasons. Even though the shape and flexibility parameters $\xi$ and $\kappa$ are constant across each cluster, the variability of scale parameter $\sigma$ (estimated locally) accounts for the high level of spatial detail of the fit. Regions with high return levels are shown in deep blue and purple colours on the map. Specific regions known to experience heavy precipitation are highlighted, such as the Cévennes, South of France \citep[with Cévenols episodes, see e.g.][]{Ducrocq2008, Vautard2015} in SON and the Canton of Ticino in southern Switzerland \citep{Isotta2014, Barton2016, Panziera2018} in SON, MAM, and JJA.

\begin{figure}[h!]
    \centering
    \includegraphics[width=12cm]{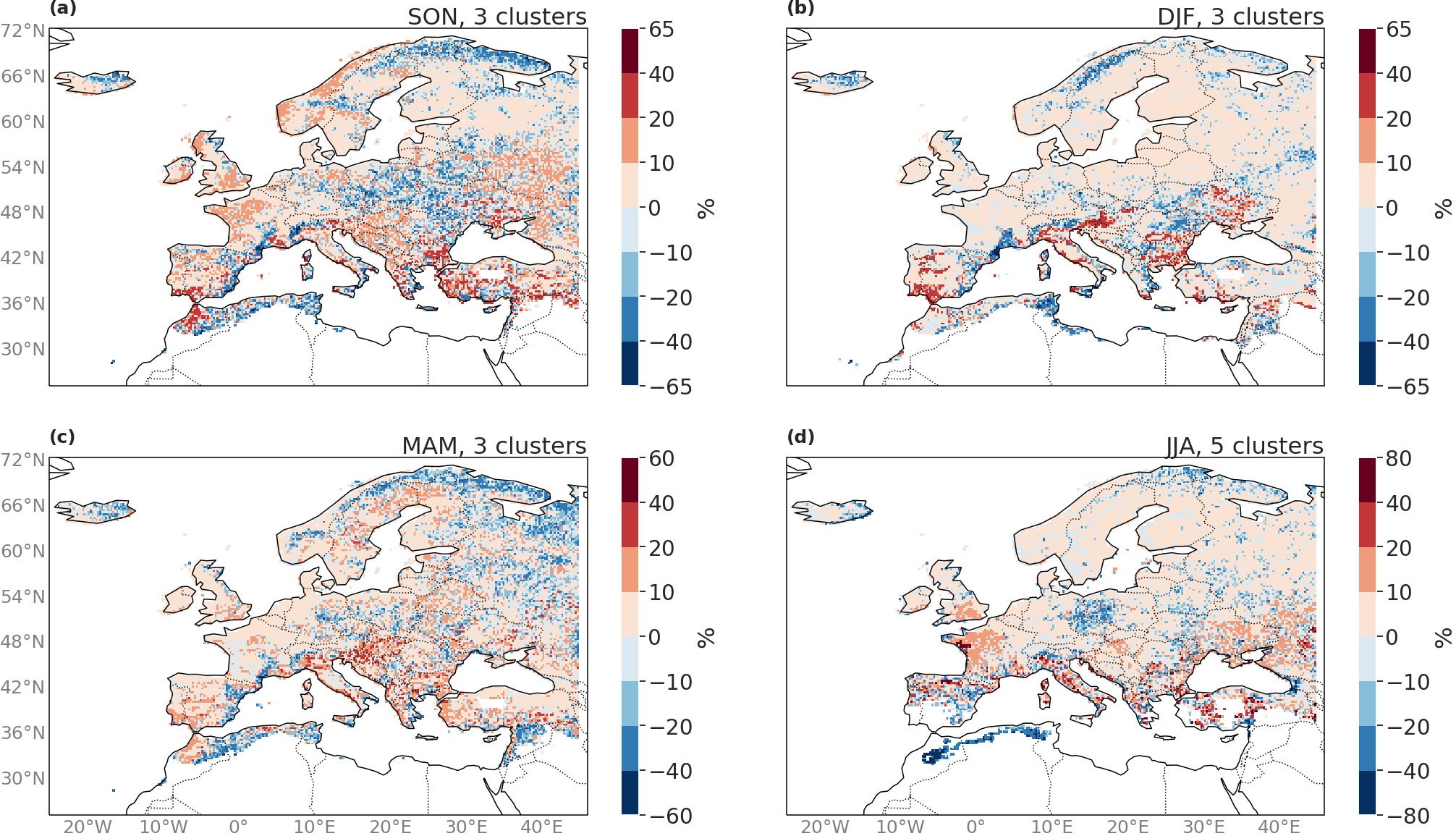}
    \caption{Relative difference between the 50-year return levels computed with the regional fitting and the local fitting.}
    \label{fig:diffRL_50}
\end{figure}

Finally, we compare the return levels obtained with the local fit and the regional fit. Figure \ref{fig:diffRL_50} displays the relative difference between the 50-year return levels for regional and local fittings. The return levels differ by less than 10\% for about 60\% of the grid points in SON and for up to 80\% of the grid points in DJF. The mean value of the absolute difference lies between 7\% (DJF) and 10\% (SON). Areas with the highest relative differences are generally located in the cluster with the highest shape parameters: those with more frequent extremes. The same maps for the 10- and 100-year return levels can be found in appendix (Fig. \ref{fig:diffRL_10} and \ref{fig:diffRL_100}).

\section{Discussion}\label{sec: discussion}

We conduct a PAM clustering procedure based on the PWM ratio $\omega$. We find that the optimal number of clusters is three in SON, DJF and MAM, and five in JJA. The higher number of clusters in JJA might be explained by a larger spatial variability of precipitation extremes in Europe in summer \citep[see e.g][in Spain]{Cortesi2014}. The same analysis was conducted for hierarchical partitioning, leading to the same results for some parametrization (results available upon request).

The choice of the optimal number of clusters is challenging. The various criteria for the choice proposed in the literature did not agree on the optimal number of clusters \citep[see e.g. ][for the use of two criteria]{pan13}. This can be explained by the large number of grid points in the analysis, resulting in more noise in the criteria than actual information about the goodness of the partitioning. In general, if many grid points are involved, we recommend using more than one criterion and checking the maps visually for the plausibility of the partition obtained.

 We analyse the impact of the number of clusters on the regional fit. For this purpose, we compare the difference between the 50-year return levels based on the PAM partition with three clusters and the one with four clusters, in SON (not shown). For a large majority of the grid points, the difference in return levels is lower than 5\%. The difference is a bit larger for a few outliers but remains lower than 25\%. The outliers are generally located in regions with a very low silhouette score for the partition with 3 clusters.

We compare our partition in Central Europe to that obtained by \cite{gvo19} over Germany, Poland, Austria and the Czech Republic. They considered extreme events between 1961 and 2013. Their approach was based on the weather extremity index. They computed Ward's linkage in the hierarchical clustering algorithm. The clusters they found exhibit a West-East pattern. The partition we obtain over these four countries also tends to separate eastern and western regions.
\cite{dar21} also found this West-East pattern in the UK. 
They delineated the regions using the most explanatory covariates (among those that were available) and then assessed their homogeneity by computing tests of \cite{hos05} on hourly precipitation.
Our results generally agree with the partition they obtained.

\begin{figure}[h!]
    \centering
    \includegraphics[width=12cm]{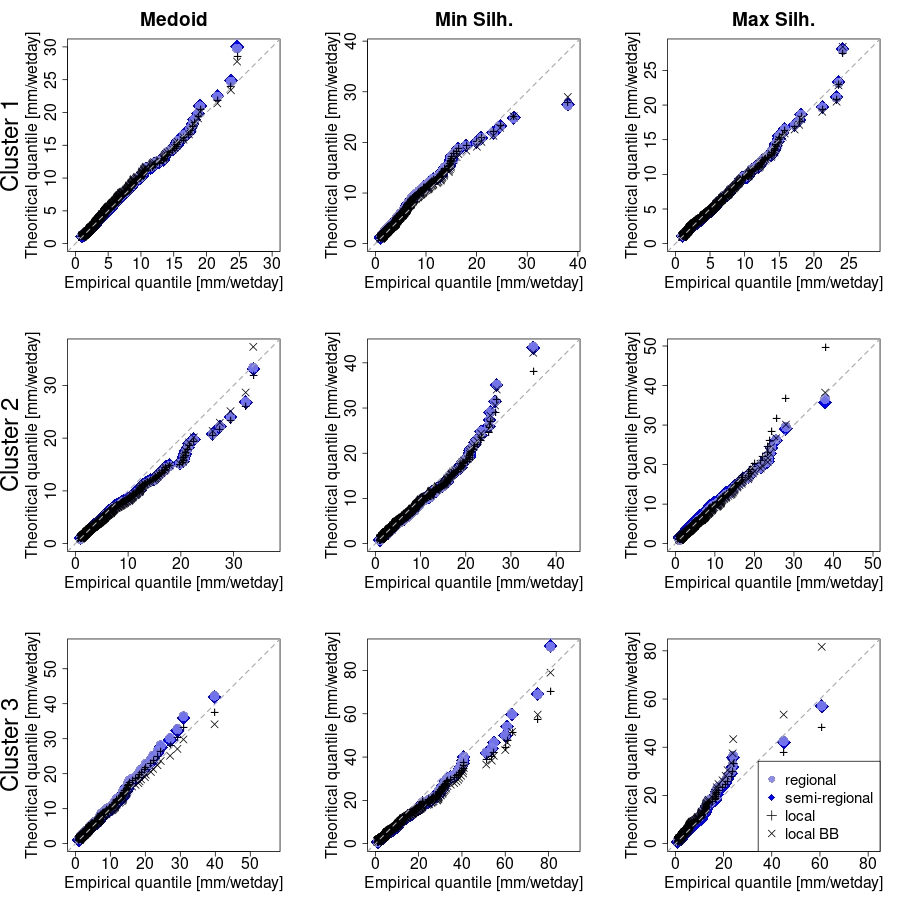}
    \caption{Example QQ plots of the regional, semiregional, local, and local Bernstein (local BB) fittings, for the medoid point (left) and the grid points with minimum (middle) and maximum (right) silhouettes in the northern (top row), intermediate (middle row), and southern (bottom row) clusters in SON (blue cluster in Figure \ref{fig:partition_PAM}).}
    \label{fig:qqplot}
\end{figure}

The regional model is more parsimonious than the local model; see Table \ref{tab: EGPD_models}. It is also more precise on well-classified points (see Figure \ref{fig:qqplot}). The semiregional and regional models have similar performance; hence, the regional model should be preferred because it is more parsimonious. 
An alternative to our fitting method would be to select only grid points with a satisfactory silhouette score (e.g. greater than 0.2) to estimate the regional parameters. The quantiles of points with very low silhouette scores would then be estimated locally. This could increase the likelihood of the fitted distribution in some cases but would also increase the number of parameters to fit. However, the performance of the regional fit was not substantially lower than the local fit for the border areas between the clusters, and the rate of rejection in the Anderson–Darling test was not substantially higher at grid points with low silhouettes. For the sake of simplicity and parsimony, we choose to keep the regional approach for all points.

The local Bernstein distributions do not seem to be substantially closer to the empirical distribution than the regional ones; see Figure \ref{fig:qqplot}.
Hence the flexibility brought by the scale parameter $\sigma$ in the regional model is sufficient to fit the data well and therefore the most parsimonious model is as precise as the others.


The spatial pattern of our seasonal 10-year return levels (Fig. \ref{fig:RL_10}) is similar to that of the yearly 10-year return levels obtained by \cite{poschlod2021} with an observational data set and the Canadian Regional Climate Model.

We also compare the return levels over Switzerland with those provided by \cite{MeteoCH} for all the seasons (see Fig. \ref{fig:Swiss_RL10}, \ref{fig:Swiss_RL50} and \ref{fig:Swiss_RL100} in appendix). This small country provides a good test case for our study because the complex orography leads to a wide variety of precipitation patterns \citep{Schmidli2002,umbricht2013, Isotta2014,  evin18}. Return levels obtained by \cite{MeteoCH} were computed by fitting a GEV to observed seasonal maxima, with a much higher spatial grid resolution than ERA-5 \citep[up to 1km, see][]{MeteoCH_spa_re}. The maps of return levels are close in terms of magnitude and exhibit very similar spatial patterns. Only the small-scale structures are not captured by ERA-5 which is due to the coarser grid resolution of ERA-5. The magnitudes of extremes are slightly underestimated in ERA-5, especially in MAM. This agrees with the study of \cite{hu2020} over Germany. They state that ERA-5 generally underestimates extremes of daily precipitation compared to observation-based gridded datasets (and weather station observations).
In our analysis, despite the regionalization (three or five clusters in Switzerland depending on the season) of two parameters out of three,  the scale parameter $\sigma$ presents sufficient variability to have correct return levels. The variability of $\sigma$ alone is sufficient to provide accurate fitting, even in a country with a complex topography and high local spatial variability of extreme precipitation.

\section{Conclusions}\label{sec: conclusion}
We derive return levels of extreme daily precipitation ($~>1$ [mm]) over Europe using regionalized parameters for the EGPD fits. The regionalization requires two steps. First, all land grid points are partitioned into a few homogeneous regions with a clustering algorithm. As distance measure, we estimate a scale-invariant ratio of PWM for each grid point, focusing on the tail of the distribution, and then use the PAM clustering algorithm to group these estimates into regions. The second step is the choice and fitting of a model to estimate return levels. We choose to fit an EGP distribution that models the full range of precipitation intensity. Only the scale parameter is allowed to vary within a homogeneous cluster, and the tail and flexibility parameters are common to all grid points in that cluster.

We assessed our regional analysis with classical statistical tools and compared it to previous analyses and return level estimates. Although parsimonious, the regional model is sufficiently flexible to capture the strong spatial variability of rainfall intensities.

This paper provides two main contributions.
We provide maps of 10-, 50- and 100-year return levels for European precipitation of ERA-5, and we have made the algorithms for clustering and regional model available in a GitHub repository\footnote{\url{https://github.com/PhilomeneLeGall/RFA_regional_EGPDk.git}}.
 
\section*{Author contributions}

Pauline Rivoire: Investigation, Conceptualization, Software, Methodology, Formal analysis, Investigation, Writing - Original Draft, Visualization. Philomène Le Gall: Investigation, Conceptualization, Software, Methodology, Formal analysis, Investigation, Writing - Original Draft, Visualization. Philippe Naveau: Supervision, Conceptualization, Methodology, Writing - Review \& Editing. Olivia Martius: Supervision, Conceptualization, Methodology, Writing - Review \& Editing. Anne-Catherine Favre: Supervision, Conceptualization, Methodology, Writing - Writing - Review \& Editing.

\section*{Acknowledgments and funding}
Within the CDP-Trajectories framework, this work is supported by the French National Research Agency in the framework of the ``Investissements d’avenir'' program (ANR-15-IDEX-02). 
P.L. and A-C.F. gratefully acknowledge financial support for this study provided by the Swiss Federal Office for Environment (FOEN), the Swiss Federal Nuclear Safety Inspectorate (ENSI), the Federal Office for Civil Protection (FOCP), and the Federal Office of Meteorology and Climatology, MeteoSwiss, through the project EXAR (``Evaluation of extreme Flooding Events within the Aare-Rhine hydrological system in Switzerland").

P.R. and O.M. acknowledge funding from the the Swiss National Science Foundation (grant number 178751).

Part of this work was supported by the  DAMOCLES-COST-ACTION on compound events, the French national program (FRAISE-LEFE/INSU and 80 PRIME CNRS-INSU), and the European H2020 XAIDA (Grant agreement ID: 101003469). 
P.N. also acknowledges the support of the French Agence Nationale de la Recherche (ANR) under reference ANR-20-CE40-0025-01 (T-REX project), and the ANR-Melody.  
 
The authors declare that they have no conflict of interest.

\section*{Data availability}
 We use ERA-5 global total daily precipitation data at resolution 0.25$^\circ$. Hourly values were downloaded from the ECMWF MARS server (a valid ECMWF account required): \url{https://apps.ecmwf.int/data-catalogues/era5/?type=fc&class=ea&stream=oper&expver=1} Forecast steps 6 to 17, variable: Total precipitation (228.128). The MARS / EMOSLIB interpolation library  has been used.

\clearpage

\bibliographystyle{ametsoc2014}
\bibliography{biblio}

\clearpage


\begin{small}
\begin{algorithm}
\caption{Regional fit of the EGPD in cluster $C$, see last row of Table \ref{tab: EGPD_models}.}\label{alg: regEGPDfitIter}
\begin{algorithmic}[1]
\State $cond=TRUE, eps=.001,  \text{ and }  u=1  \text{(mm) }$ 
\Procedure{Input}{Rainfall Matrix  for  cluster $C$} 
\State Remove dry days by only taking $\{Y(s)|Y(s)>u\}$
\State Fit  local Model  at each location $s \in C$ 
\State Denote  $\kappa_0$ and $\xi_0$ the cluster  means  of $\kappa$ and $\xi$  from Step 4 
\State Compute $m(s)$ the sample mean  at   each  $s\in C$
\While{$cond=TRUE$}
\State 
\small{ Compute $$\sigma_{new}(s) =\dfrac{{\xi}_0~ m(s)}{\frac{{\kappa}_{0}}{{\Bar{F}}(u)}IB\left(H_{{\xi}_0}\left(\frac{u}{{\sigma}_{0}}\right),1,{\kappa}_{0},1-\xi_0\right)-1}$$
\indent \indent  where $IB(.,.,.)$ is the incomplete Beta function}

\State \textbf{if} $|{\sigma}_{new}-{\sigma}_{0}|< eps$ \textbf{then} 
\State $cond=FALSE$
\State \textbf{end if}
\State \indent $\sigma_{0}\leftarrow \sigma_{new}$
\EndWhile
\State \textbf{end while}
\State Return $(\kappa_0,\sigma_{0},\xi_0)^T$

\EndProcedure
\State \textbf{end procedure}

\end{algorithmic}
\end{algorithm}
\end{small}

\clearpage



\begin{table}[!ht]
\hskip-1cm
    \centering
\small
    \begin{tabular}{|cccc|}
    \hline
    &&&\\
       \large{Models}  &  \large{Flexibility function $G$} & \large{$\xi$} & \large{$\sigma$} \\
       \hline  
        &&&\\

        Local Bernstein  & $G_i =$ Bernstein polynomials, site specific & site specific & site specific \\ 
      &&&\\
     
        \textbf{Local}   & $G_i(u) = u^{\kappa_i}, \kappa_i > 0$ site-specific & site-specific & site-specific \\ 
&&&\\
        \textbf{Semiregional} & $G_i(u) = u^{\kappa_i}, \kappa_i > 0$ site-specific & constant on each cluster & site-specific \\
        &&&\\
        \textbf{Regional} & $ G_i(u) = u^\kappa, \kappa>0$ constant on each cluster & constant on each cluster & site-specific\\
       
        &&&\\
        \hline
        
    \end{tabular}
    \caption{Description of the four EGPD models, with various complexity compared in Section \ref{sec: results}\ref{sec: assess_fit_res}. The Bernstein EGPD  is presented in \cite{tencaliec20}. The local EGPD is introduced in \cite{nav16} and its regional version in \cite{legall21}. The comparison is mainly conducted between the local, the semiregional and the regional fitting (in bold).}
    \label{tab: EGPD_models}
\end{table}


\clearpage


\begin{table}[!ht]
    \centering
    \begin{tabular}{l c c c c}
       \textrm{Model} & \textrm{SON} & \textrm{DJF} & \textrm{MAM} & \textrm{JJA}\\
       \hline
       \textrm{local} & 91\% & 89\% & 90\% & 87\% \\
       \textrm{semiregional} & 89\% & 87\% & 88\% & 83\% \\
       \textrm{regional} & 88\% & 88\% & 88\% & 84\%
    \end{tabular}
    \caption{Anderson--Darling test at a risk level of 5\%: Percentages of grid points for which the hypothesis of equality between the empirical distribution and the fitted distribution is not rejected. Distributions are fitted locally semiregionally and regionally; see second, third and last row of Table \ref{tab: EGPD_models}.}
    \label{tab:AD_results}
\end{table}


\clearpage


\begin{table}[!ht]
    \centering
    \begin{tabular}{l c c c c}
       \textrm{Model} & \textrm{SON} & \textrm{DJF} & \textrm{MAM} & \textrm{JJA} \\
       \hline
       \textrm{local} & -30,888 & -27,984 & -30,016 & -43,704\\
       \textrm{semiregional} & -69,792 & -67,614 & -69,138 & -79,400\\
       \textrm{regional} & -108,702 & -107,250 & -108,266 & -115,106\\
    \end{tabular}
    \caption{Akaike information criterion over Europe for each model and season.}
    \label{tab:AIC_results}
\end{table}


\clearpage


\clearpage

\clearpage

\clearpage
\clearpage

\clearpage

\clearpage

\appendix

\begin{appendices}

\counterwithin{figure}{section}

\section{Difference between the regional and the local fittings}

\begin{figure}[h!]
    \centering
    \includegraphics[width=12cm]{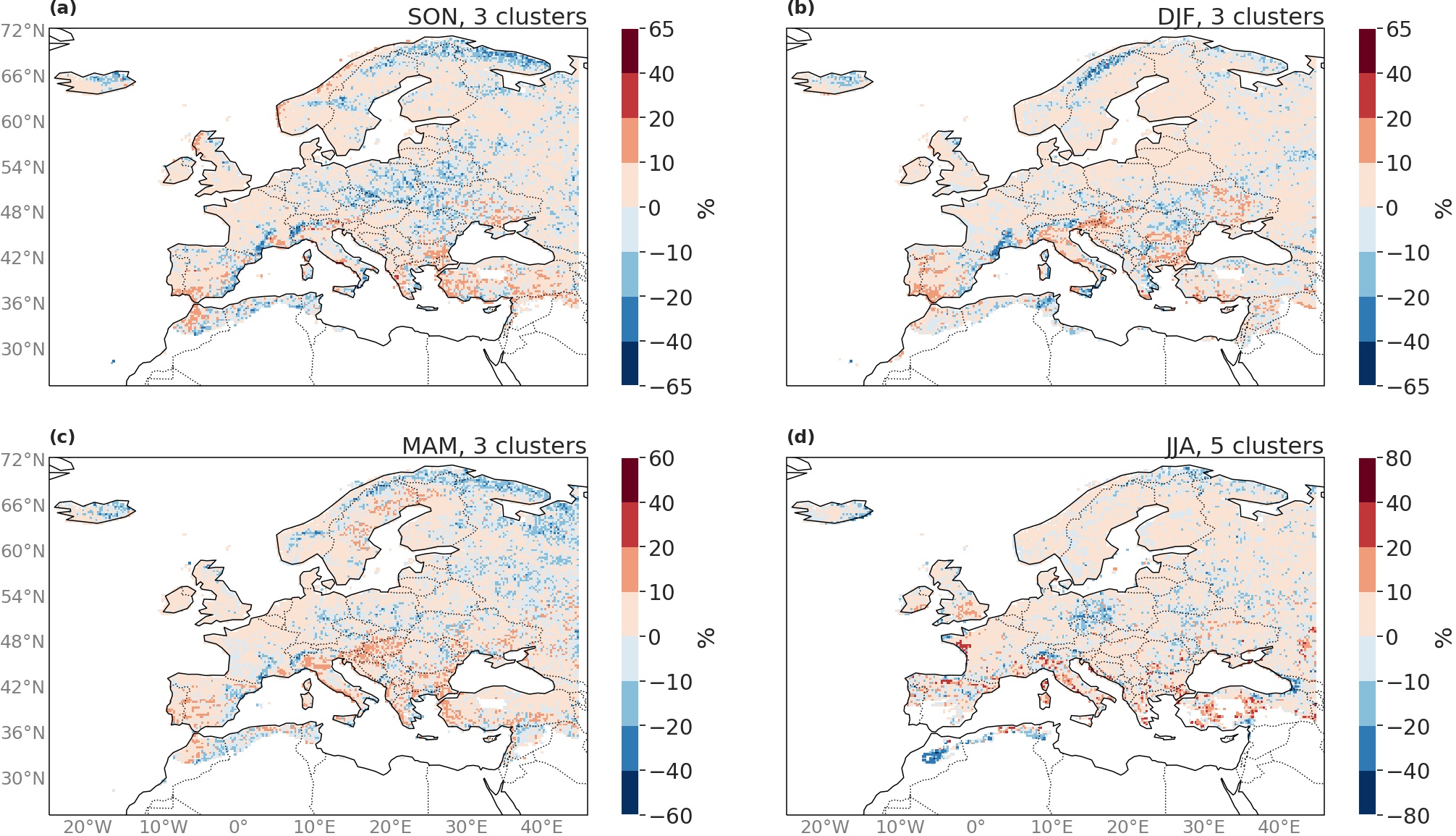}
    \caption{Relative difference between the 10-year return levels computed with the regional fitting and the local fitting.}
    \label{fig:diffRL_10}
\end{figure}

\clearpage

\begin{figure}[h!]
    \centering
    \includegraphics[width=12cm]{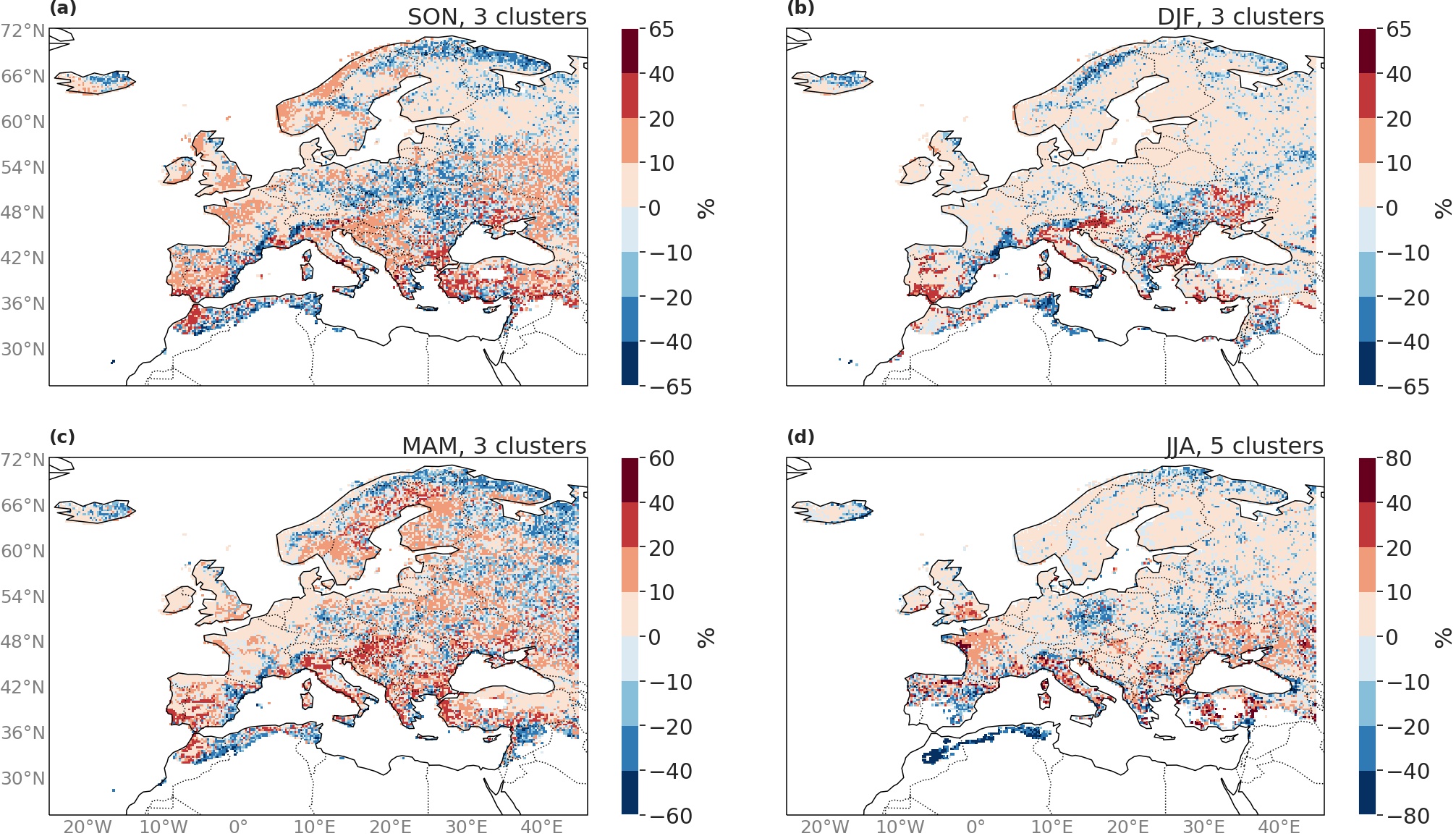}
    \caption{Relative difference between the 100-year return levels computed with the regional fitting and the local fitting.}
    \label{fig:diffRL_100}
\end{figure}

\clearpage

\section{Retun levels in Switzerland}

\begin{figure}[h!]
    \centering
        \includegraphics[width=12cm]{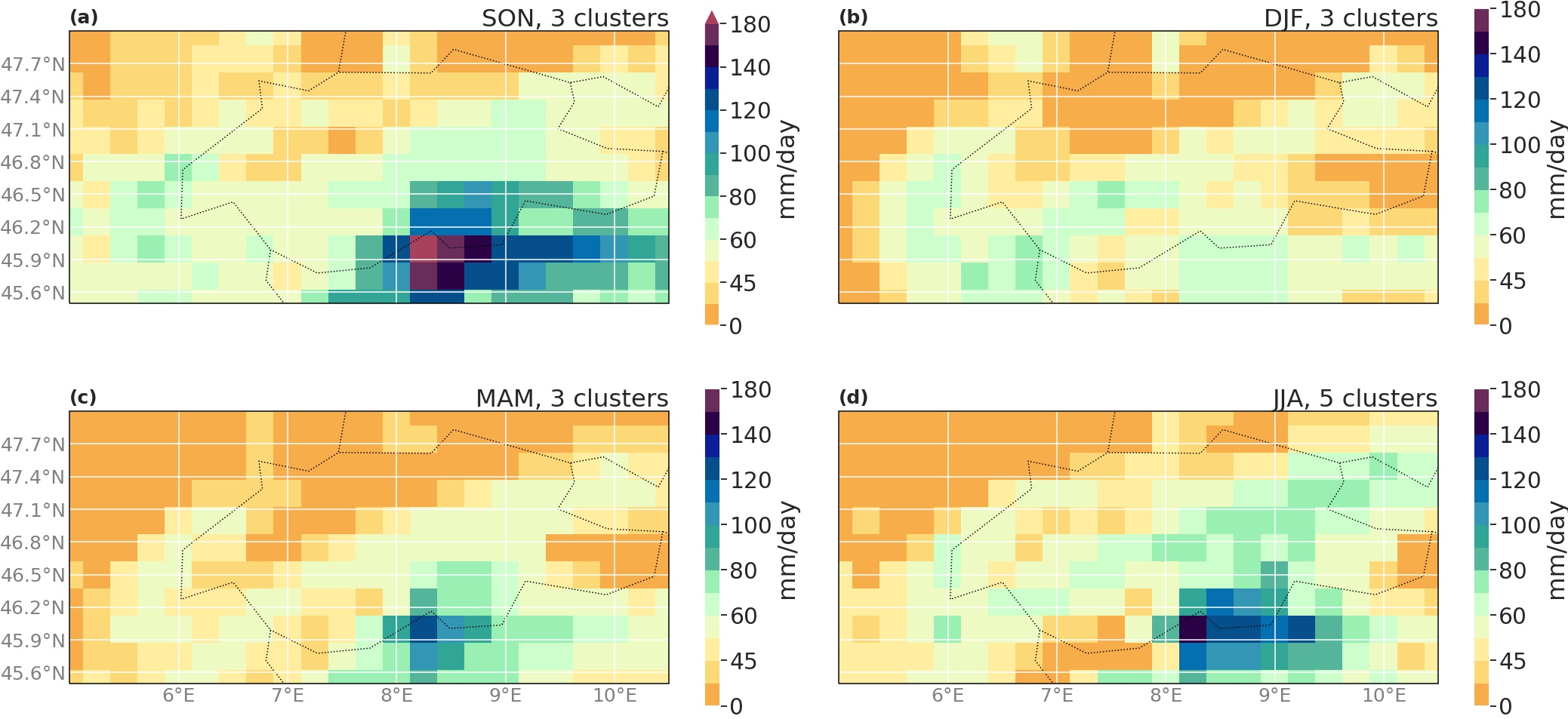}
        \caption{10-year return levels in Switzerland computed with the regional fitting, for a comparison with the one provided by \cite{MeteoCH}.}
    \label{fig:Swiss_RL10}
\end{figure}

\clearpage

\begin{figure}[h!]
    \centering
        \includegraphics[width=12cm]{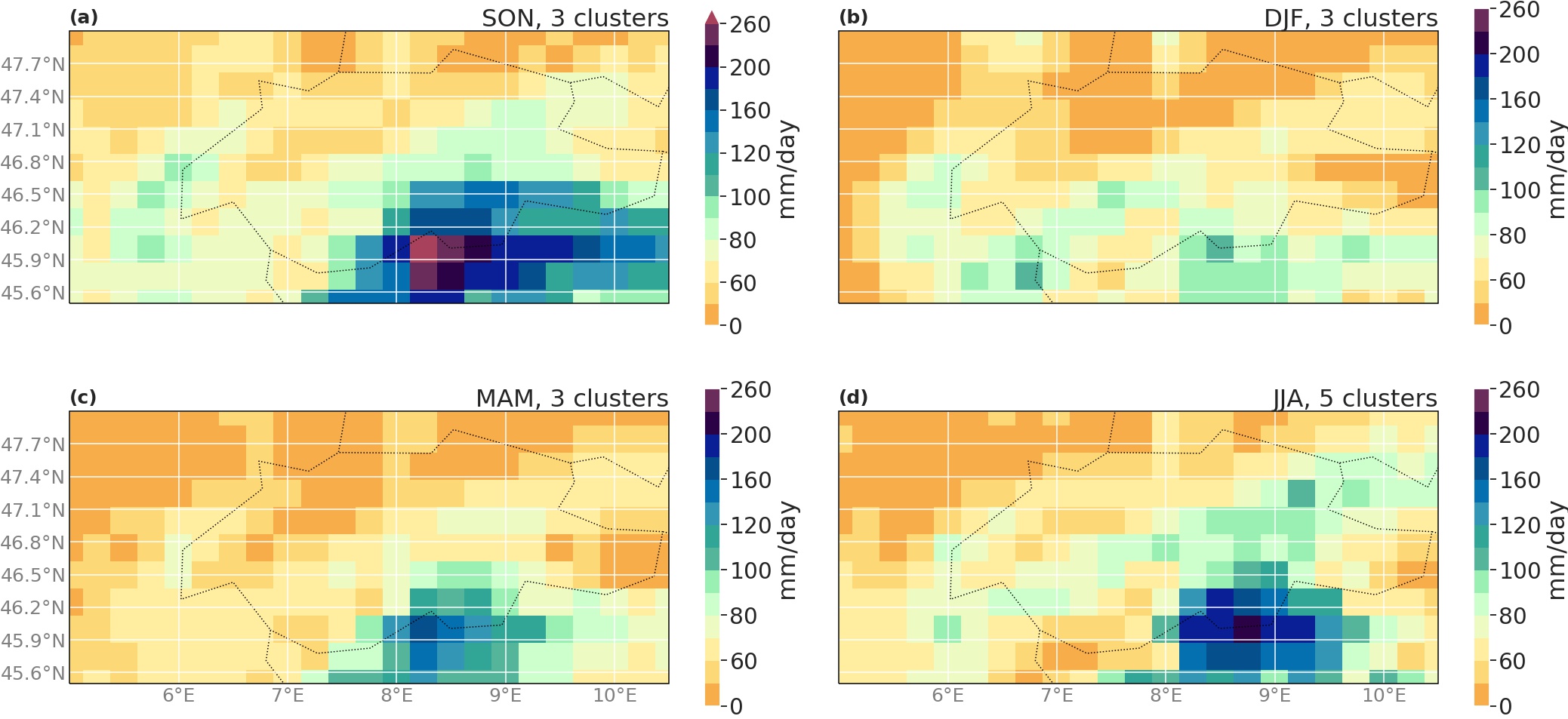}
        \caption{50-year return levels in Switzerland computed with the regional fitting, for a comparison with the one provided by \cite{MeteoCH}.}
    \label{fig:Swiss_RL50}
\end{figure}

\clearpage

\begin{figure}[h!]
    \centering
        \includegraphics[width=12cm]{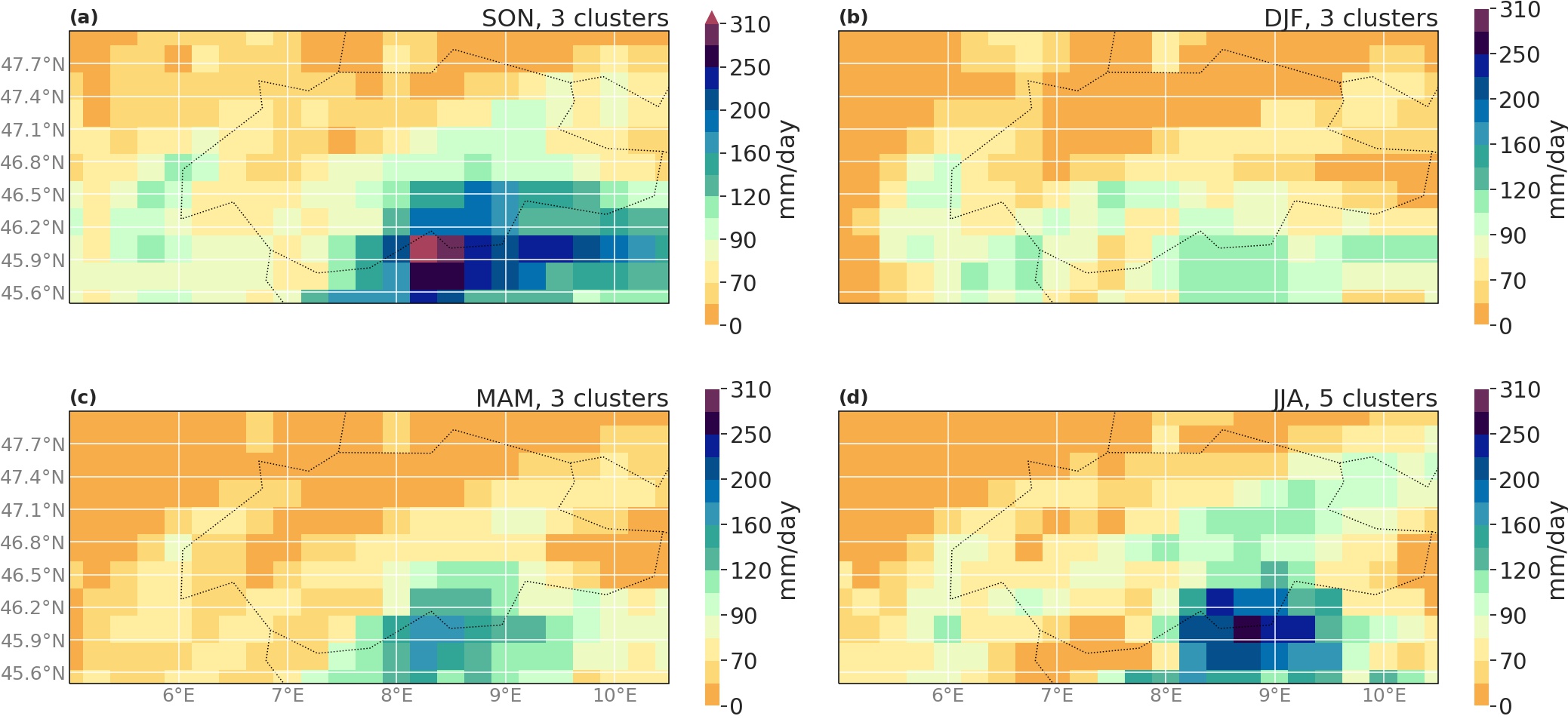}
        \caption{100-year return levels in Switzerland computed with the regional fitting, for a comparison with the one provided by \cite{MeteoCH}.}
    \label{fig:Swiss_RL100}
\end{figure}

\clearpage

\end{appendices}

\end{document}